# Software Architecture for Dynamic Adaptation of Heterogeneous Applications

Christine Louberry
LIUPPA – IUT Bayonne
2 Allée Parc Montaury
64600 Anglet
+33559574326

Marc Dalmau
LIUPPA – IUT Bayonne
2 Allée Parc Montaury
64600 Anglet
+33559574321

Philippe Roose
LIUPPA – IUT Bayonne
2 Allée Parc Montaury
64600 Anglet
+33559574348

{christine.louberry ; marc.dalmau ; philippe.roose}@iutbayonne.univ-pau.fr

## ABSTRACT

The recent apparition of mobile wireless sensor aware to their physical environment and able to process information must allow proposing applications able to take into account their physical context and to react according to the changes of the environment. It suppose to design applications integrating both software and hardware components able to communicate. Applications must use context information from components to measure the quality of the proposed services in order to adapt them in real time.

This work is interested in the integration of sensors in distributed applications. It present a service oriented software architecture allowing to manage and to reconfigure applications in heterogeneous environment where entities of different nature collaborate: software components and wireless sensors.

## Categories and Subject Descriptors

D.3.3 [**Software Engineering**]: Software Architectures – *domain-specific architectures.*

## 1. INTRODUCTION

Depuis quelques années, l'émergence des capteurs sans-fil a engendré de nombreux défis dans les domaines des réseaux et des architectures logicielles. La grande majorité des travaux liés aux réseaux de capteurs concernent l'optimisation des ressources matérielles (capacité de calcul, énergie) et réseaux (contrôle de congestion, agrégation des données, etc.). Les capteurs y sont utilisés pour leurs fonctions propres de mesure de l'environnement et leur capacité à transmettre et relayer l'information, en veillant à en maximiser leur durée de vie. Il existe actuellement peu de recherches sur l'intégration des capteurs dans des environnements hétérogènes où collaborent composants logiciels et capteurs. Cet article s'intéresse particulièrement au domaine des architectures logicielles et à l'intégration des capteurs dans des applications distribuées.

Face à la demande grandissante pour des services de plus en plus riches et personnalisés, le défi aujourd'hui est de proposer des applications qui s'adaptent tant aux souhaits de l'utilisateur qu'à l'environnement réel. Les capteurs peuvent permettre de proposer des applications conscientes de leur contexte physique et réactives aux évolutions de l'environnement. L'exemple que nous avons choisi pour illustrer l'intégration de capteurs dans des applications distribuées est celui d'une application de télésurveillance. Des capteurs infrarouges et des capteurs équipés d'une micro-caméra sont disséminés sur un secteur qui doit être surveillé. Les capteurs infrarouges sont capables de détecter des intrusions dans le secteur. La détection d'un intrus entraîne le démarrage des caméras voisines. La collaboration des caméras avec un composant logiciel d'analyse vidéo permet de déterminer la trajectoire probable de l'intrus et par conséquent de démarrer les prochaines caméras susceptibles de se trouver sur cette trajectoire ou d'orienter les caméras mobiles pour obtenir des images sous différents angles de vue. Si un capteur infrarouge vient à tomber en panne, il est possible d'utiliser le capteur vidéo le plus proche en tant que capteur de détection en reconfigurant l'application en ce sens. Bien entendu, la qualité du service peut être dégradée, lorsqu'il n'y a pas assez de lumière par exemple. Si ses ressources le permettent ce capteur pourrait même réaliser les deux fonctionnalités, détection et suivi, simultanément. L'objectif de nos travaux est de proposer une architecture permettant de détecter automatiquement ce type de situations et de réorganiser les divers composants de l'application de façon à continuer à assurer le service malgré tout. Nous utilisons la définition de qualité de service essentiellement dans le sens où l'application doit assurer le service qu'elle propose quelle que soit l'évolution de son environnement. En effet, un des objectifs de cette architecture est de faire en sorte que les choix de configurations doivent permettre de maximiser la durée de vie des capteurs et par conséquent de maximiser la durée de vie de l'application.

Dans cet article, nous pensons qu'il est intéressant de ne plus considérer les capteurs simplement comme des dispositifs capables d'effectuer des mesures. En effet, lorsqu'ils n'utilisent pas la totalité de leur mémoire et de leur capacité de calcul, il est possible d'y héberger d'autres composants logiciels en relation ou non avec la fonction du capteur. Ainsi les capteurs peuvent participer à l'infrastructure matérielle des applications classiques en offrant de nouvelles possibilités pour héberger des fonctionnalités. Ceci permet de proposer de nouvelles configurations pour les applications et, par conséquent, d'accroître les possibilités d'offrir la meilleure qualité de service possible.

Le paradigme orienté service est aujourd'hui fréquemment utilisé pour le développement d'applications largement distribuées devant offrir une forte interopérabilité et évolutivité. Tout comme le paradigme composant, il favorise la réutilisation et la maintenance des applications mais il offre également des mécanismes pour la description, la publication et la découverte des services. De plus, un service est une entité autonome et indépendante de la plateforme qui le supporte ce qui permet des reconfigurations aisées dans un environnement hétérogène. Enfin le mode de communication qu'ils utilisent permet une plus grande transparence. En effet, lorsque deux composants doivent dialoguer, il est nécessaire qu'ils se connaissent avant l'exécution alors que les services se lient à la volée, pendant l'exécution grâce aux mécanismes de découverte et de publication [3] [10]. Pour toutes ces raisons et dans le but de gérer la qualité de service des applications distribuées dans des environnements hétérogènes, nous avons choisi de proposer une architecture à base de services.

La section 2 présente l'architecture proposée pour superviser et reconfigurer des applications hétérogènes distribuées. Elle détaille les composants et les services utilisés ainsi que les choix de déploiement selon les hôtes utilisés. La section 3 présente les travaux apparentés sur les architectures logicielles pour la reconfiguration d'applications utilisant des capteurs. La section 4 conclut cet article et présente les perspectives de ces travaux.

## 2. ARCHITECTURE DES APPLICATIONS INTEGRANT DES CAPTEURS

Ces travaux s'intéressent à l'intégration de capteurs dans les applications distribuées. Cependant la conception de telles applications peut devenir complexe si l'on doit prendre en compte le caractère physique ou logiciel des composants. Les recherches décrites dans [2] proposent un modèle de composant pour la reconfiguration dynamique d'applications multimédias distribuées, Osagaia. Ce modèle met en avant la séparation des préoccupations. En effet, il sépare la partie métier du composant (partie fonctionnelle) de la partie qui gère les propriétés non-fonctionnelles comme les échanges de données par exemple. Nous avons fait le choix de modéliser tous les composants de la même façon : une fonctionnalité -la partie métier- encapsulée dans un gestionnaire -un conteneur- constituant la partie non-fonctionnelle. Nous avons adopté ce modèle pour concevoir les applications constituées de composants logiciels et de capteurs. Nous l'avons adapté aux capteurs de façon à faire abstraction de la nature des composants et permettre une conception unifiée et simplifiée [8].

### 2.1 Les composants

Chaque capteur ou composant logiciel offre une fonctionnalité particulière appelée Composant Métier (CM) qui contient l'implémentation d'un traitement particulier de données. Les capteurs, quant à eux, offrent en outre des fonctionnalités particulières de mesure de l'environnement qu'ils sont seuls à proposer (température, luminosité, déplacement).

Le CM est contenu dans un Processeur Elémentaire (PE) (Fig. 1). Un PE est un conteneur qui assure les échanges de données avec les autres composants grâce à deux unités d'échange : l'unité d'entrée (UE) et l'unité de sortie (US). Il contrôle également le fonctionnement du CM grâce à une unité de contrôle (UC). Cette UC constitue le lien avec le service de supervision, elle lui envoie l'état du CM et en interprète les commandes. Un PE encapsule un seul CM et par conséquent gère uniquement l'état de son CM.

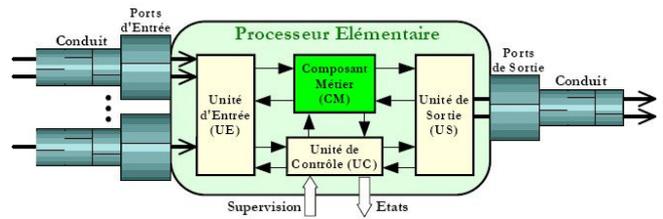

**Figure 1. Architecture interne d'un PE.**

Les PEs sont interconnectés par des Conduits. Un Conduit est une entité qui fournit des fonctionnalités de connexion et de transport d'information. Son architecture est sensiblement identique à celle d'un PE : il possède des unités d'échanges et une unité de contrôle pour superviser les transferts de données (Fig. 2). Pour se connecter, les PEs et les Conduits possèdent des ports d'entrée et de sortie. La connexion se fait de la façon suivante : le port de sortie d'un PE est connecté au port d'entrée d'un Conduit et inversement le port de sortie d'un Conduit est connecté à un des ports d'entrée d'un PE. Un Conduit peut transporter un ou plusieurs flux et proposer des politiques de transport dédiées (synchronisation, etc.).

L'un des intérêts de nos travaux est de gérer la qualité de service (QdS) dans les environnements hétérogènes. Les applications sont supervisées par une plateforme qui évalue la QdS et opère des reconfigurations quand le niveau se dégrade ou lorsqu'une amélioration est possible [7]. Lors de la conception, des familles de configurations sont élaborées selon des niveaux de QdS puis stockées sur la plateforme sous la forme de graphes [7]. Lorsqu'une reconfiguration est nécessaire, celle-ci recherche parmi les configurations possibles celle qui permettra d'obtenir une QdS optimale et la met en œuvre. Cette plateforme est modélisée sous forme de services.

### 2.2 Les services

Les architectures à services sont souvent utilisées pour les systèmes ayant besoin d'être dynamiquement reconfigurés. Nos systèmes étant appelés, eux aussi, à être reconfigurés fréquemment pour garantir la meilleure qualité de service possible, nous avons fait le choix de concevoir une architecture à services. Elle est composée de quatre services : *Supervision*,

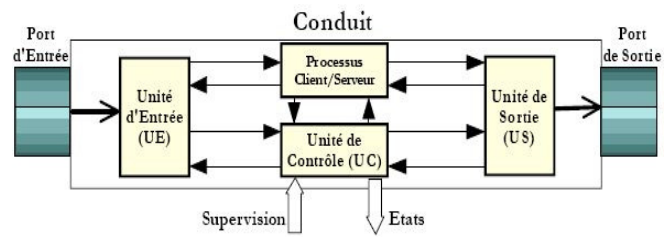

**Figure 2. Architecture interne d'un Conduit.**

*Usine à Conteneur*, *Usine à Conduit* et *Routage*.

Le service *Supervision* est le service principal de la plateforme. Son rôle est de surveiller le fonctionnement de l'application. Pour cela il reçoit des différents composants de l'application des informations d'état à partir desquelles il évalue la qualité du service rendu. A l'issue de cette évaluation, il décidera de reconfigurer ou non l'application. Il a la charge d'envoyer aux services *Usine à Conteneur* et *Usine à Conduit*, qui seront décrits plus loin, les directives pour qu'ils créent les conteneurs et les composants de liaison adaptés à la nouvelle configuration et offrant le niveau de qualité de service requis. Pour évaluer au mieux la qualité de service et décider de la meilleure configuration possible, le service *Supervision* doit savoir quels sont les composants qu'il peut atteindre et, par conséquent, quelles sont les routes valides pour les atteindre. Il peut obtenir ces renseignements auprès du service *Routage* qui sera également décrit plus loin.

Le service *Usine à Conteneur* est un service permettant de construire un conteneur adapté au composant métier qu'il doit encapsuler. En effet, le conteneur n'est pas le même s'il doit encapsuler un composant logiciel situé sur un site non contraint, qu'on appellera site fixe (site dont les ressources ne sont pas critiques, par exemple un ordinateur de bureau) ou sur un site léger (site dont les ressources sont restreintes comme un capteur sans-fil, un téléphone mobile, un PDA, etc.) Le conteneur est également différent selon que le composant métier qu'il encapsule utilise un mécanisme de communication par évènement ou un mécanisme de communication par appel de méthode ou encore une boîte à lettres. Les CM disponibles sont situés dans un entrepôt. Le service *Usine à Conteneur* doit donc se connecter à l'entrepôt pour charger le CM demandé par le service *Supervision*. Ensuite il adapte l'interface de communication au mécanisme utilisé par le CM et déploie le CM et le PE adapté. Lorsque survient une reconfiguration, le service *Supervision* lui indique le CM et par conséquent le conteneur qu'il doit supprimer.

Le service *Usine à Conduit* est un service permettant de construire tout type de conduit adapté à l'application : transmission synchrone des données ou non, transmission en temps réel ou non, etc. Lorsqu'une reconfiguration est requise, ce service est averti par le service *Supervision* de la nécessité de relier des composants métier par un conduit. Il obtient de la part du service *Supervision*, la localisation du composant source et celle du composant cible. Il obtient également les informations de contraintes de transmission liées à l'application : synchrone, asynchrone, temps réel ou non, etc. A partir de ces informations, il construit le conduit adapté et connecte chacune de ses extrémités aux composants à relier. Les informations de reconfiguration envoyées par le service *Supervision* peuvent également provoquer des changements de points de connexion. Ces changements seront effectués par la suppression du conduit jusqu'alors utilisé et la construction d'un nouveau conduit de source et/ou de cible différentes.

Le service *Routage* est un service permettant de conserver une table de routes à jour pour atteindre tous les composants de l'application. Il est à la fois un service espion et un service d'information. Il est un service espion lorsqu'il détecte la disparition d'une route en cours d'utilisation. Dans ce cas, il avertit par un message prioritaire le service *Supervision* qu'il est nécessaire et urgent d'effectuer une reconfiguration pour trouver une autre route et continuer d'assurer le fonctionnement de l'application. Il assure également un service d'information lorsqu'il détecte un changement de route, en avertissant le service *Supervision* qui décidera, selon la route concernée d'évaluer s'il est opportun de reconfigurer l'application.

Comme il est introduit dans la section 1, les applications utilisent différents types de dispositifs pour supporter leurs composants (mobile, non mobile, ressources contraintes ou non, etc.). Par conséquent deux catégories d'hôtes (ou sites) ont été définies. Les capteurs et les périphériques mobiles sont qualifiés d'hôtes légers en raison de leurs contraintes de ressources. A l'inverse d'un hôte fixe comme un PC, un hôte léger a de fortes contraintes d'énergie, de fortes limitations de capacité de calcul et de réseau, etc. Les paragraphes suivants décrivent pour chaque catégorie d'hôte, la démarche de déploiement des composants.

## 2.3 Cas d'un hôte fixe

La plateforme de supervision, au même titre que l'application, est distribuée sur tous les hôtes participant à la configuration. Chaque hôte héberge une partie de la plateforme mais tous les hôtes ne peuvent pas supporter tous les services. Nous considérons les hôtes fixes, tels les PCs, comme ayant des contraintes de ressources négligeables face à celles des hôtes de plus en plus légers comme les PDA, les téléphones mobiles, les capteurs, etc. Les hôtes fixes sont capables d'héberger les quatre services précédemment décrits et de les faire fonctionner (Fig. 3). Pour les mêmes raisons, les conteneurs et les conduits peuvent être déployés dans leur totalité sur un hôte fixe.

Le service *Supervision* avec l'aide du service *Routage* a connaissance à tout instant de la configuration de l'application. A chaque reconfiguration, il envoie les informations nécessaires aux services d'*Usine à Conteneur* et d'*Usine à Conduit* pour créer et/ou supprimer des PEs et des conduits.

Quand il déploie un PE, le service *Usine à Conteneur* le déploie avec ses unités d'échange et son UC. De la même façon, le service *Usine à Conduit*, déploie sur l'hôte fixe source l'extrémité du conduit contenant le port d'entrée avec une partie de l'UC et sur l'hôte cible l'extrémité contenant le port de sortie avec une partie de l'UC (ou tout sur le même hôte s'il s'agit d'une connexion interne). Dans le paragraphe suivant, nous verrons que le déploiement des PEs et des Conduits est différent pour un hôte léger.

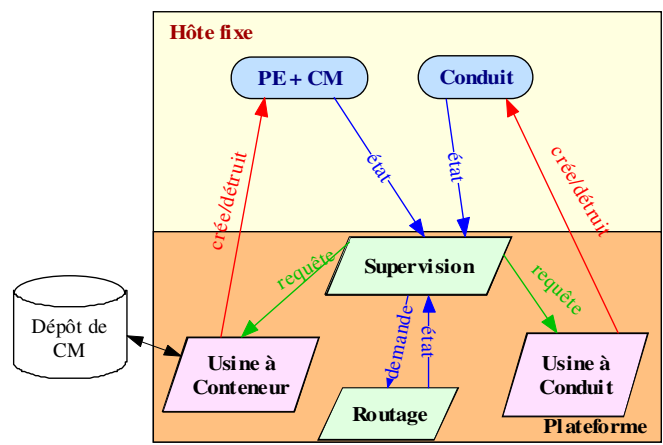

**Figure 3. Déploiement de la plateforme sur un hôte fixe**

## 2.4 Cas d'un hôte léger

Un hôte léger est limité en termes d'énergie, de capacité de calcul et de réseau. Par conséquent, il n'est pas souhaitable de le surcharger en hébergeant la plateforme dans sa totalité comme le fait un hôte fixe. Une sélection des services indispensables a été faite. Ce sont ces services qui seront effectivement hébergé sur l'hôte léger. Les autres seront déportés sur l'hôte fixe le plus proche. Ainsi le service *Routage* doit impérativement se trouver sur l'hôte léger sinon il n'a aucun moyen de connaître tous les composants (et par conséquent tous les hôtes) qu'il peut atteindre. En revanche, la supervision des hôtes légers sera assurée par le service *Supervision* de l'hôte fixe le plus proche, appelé hôte correspondant (Fig. 4). Le service *Usine à Conteneur* est dirigé par le service *Supervision*, il sera de la même façon situé sur l'hôte correspondant. A chaque reconfiguration, le service *Supervision* lui indique quel CM il doit encapsuler et la ou les politiques de communication qu'il doit implanter pour garantir le bon fonctionnement du composant. Une fois que le conteneur est créé il peut être déployé. Enfin le service *Usine à Conduit* qui est lui aussi dirigé par le service *Supervision* sera également déporté sur l'hôte correspondant. Il reçoit les informations sur les Conteneur qu'il doit créer ainsi que les contraintes qu'ils doivent respecter : synchronisation, etc. Ensuite les extrémités du conduit sont déployées sur l'hôte de chacun des conteneurs à relier.

La démarche de déploiement des PE et des Conduits est différente de celle des hôtes fixes. Pour éviter de surcharger les hôtes légers, les UC des PE sont déportées sur l'hôte correspondant de minimiser les transmissions réseaux qui consomment beaucoup d'énergie sur les capteurs. L'UC n'est sollicitée que ponctuellement par le service *Supervision* contrairement aux unités d'échange indispensables pour les communications avec les autres composants de l'application. C'est pourquoi, nous jugeons que déporter l'UC sur l'hôte correspondant n'aura pas de lourdes conséquences sur la consommation d'énergie des hôtes légers mais permettra de libérer des ressources de mémoire et de processeur. De la même façon, les UC des conduits sont déportées sur l'hôte correspondant. La table 1 récapitule le déploiement des conduits et des PE encapsulant les composants métiers. Les parties indispensables sont supportées par l'hôte léger et les autres par l'hôte correspondant.

**Table 1. Déploiement de composants sur un hôte léger et un hôte fixe**

|  | Conduit | PE + CM |
|---|---|---|
| Hôte léger | Extrémité + une partie de l'UC | CM + UE + US + une partie de l'UC |
| Hôte correspondant | UC | UC |

|  | Conduit | PE + CM |
|---|---|---|
| Hôte fixe | Extrémité + UC | CM + UE + US + UC |

Cependant les hôtes légers tels les capteurs, les PDA et les téléphones mobiles peuvent se déplacer et ainsi perdre la liaison réseau avec leur hôte correspondant. Dans ce cas, le service *Routage* va alerter le service *Supervision* que la topologie a été modifiée. Ce dernier va rechercher un nouvel hôte correspondant et ordonner d'y déplacer les UC orphelines. Les auteurs de [13] proposent un tel mécanisme de migration pour les capteurs sans-fil. Ce mécanisme ne peut être utilisé que dans le cas où les capteurs sont identiques, c'est-à-dire qu'ils disposent exactement des mêmes composants. Si la configuration choisie pour l'application prévoit des capteurs identiques, la migration proposée dans [13] pourra être mise en œuvre par la plateforme.

Dans la pratique, les capteurs font partie d'une catégorie particulière d'hôte léger, très contraint, que nous comparerons au standard CLDC[1] de la machine virtuelle Java. Les capteurs ont la particularité de ne pas contenir les outils nécessaires pour le chargement dynamique de composant. En effet, les différents programmes que l'utilisateur souhaite exécuter sur un capteur doivent être chargés sous forme de paquetages. Chaque fois qu'un composant doit être ajouté, il faut créer un nouveau paquetage comportant les composants actuels ainsi que le nouveau composant et ensuite le charger sur le capteur.

Nous proposons donc un déploiement spécifique de la plateforme pour les capteurs. En effet, puisque ni les conteneurs, ni les conduits ne peuvent être déployés à la volée, nous proposons que tous les services de la plateforme soient hébergés sur le capteur. Cependant, le service *Usine à Conteneur* est réduit à ne produire que des conteneurs spécifiques au capteur. De la même façon, le service *Usine à Conduit* ne produit que des Conduits spécifiques aux contraintes du capteur. Enfin, le paquetage chargé sur le capteur comprendra également le dépôt des CM qui pourront être mis en œuvre.

## 3. IMPLÉMENTATION

Le modèle de composant unifié Osagaia a été conformément implémenté en Java comme nous l'avons proposé dans la section

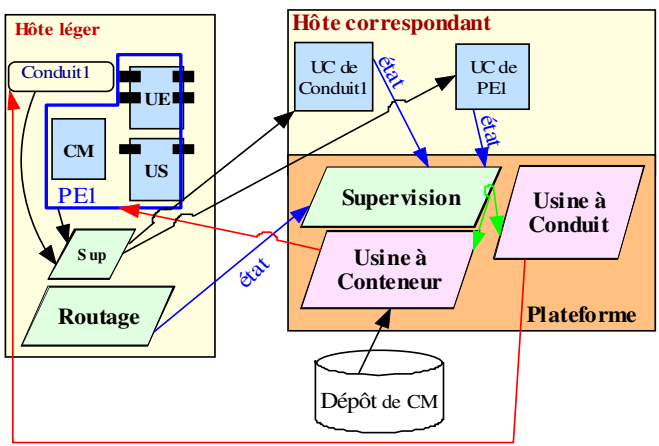

**Figure 4. Déploiement de la plateforme sur un hôte léger.**

---

[1] Connected Limited Configuration Device

2.1 et [8]. Cette implémentation, n'incluant pas la plateforme, a donné lieu à l'élaboration d'un prototype de vidéoconférence permettant une adaptation dynamique au contexte et mettant en œuvre des conduits proposant une politique de synchronisation des flux de données. L'archive est disponible à l'url suivante : http://www.iutbayonne.univ-pau.fr/~roose/pub/recherche/osagaiakorrontea.

La plateforme de reconfiguration quant à elle (Fig. 5), a donné lieu également à l'implémentation d'un simulateur sous Labview disponible à l'url suivante : http://www.iutbayonne.univ-pau.fr/~roose/pub/recherche/kalinahia. Ce simulateur permet de montrer que la plateforme, après plusieurs choix de configuration, converge vers une configuration stable de l'application permettant d'obtenir une QdS tendant vers l'optimum (ce dernier ne pouvant être atteint a priori puisqu'il est np-complet).

Les différentes implémentations précédemment citées, bien qu'elles soient complètes d'un point de vue fonctionnel, ne peuvent pas être déployées ex-nihilo sur des capteurs, beaucoup trop contraints matériellement. Nous allons donc proposer une adaptation de la plateforme, particulièrement allégée pour les périphériques fortement contraints. Notre choix de capteurs s'est porté sur des Sun Spot de Sun Microsystems. Ce choix est guidé par leur capacité à implémenter une machine virtuelle Java compatible CLDC 1.1, Squawk [13].

Il est à noter que nos travaux sont supportés par Sun Microsystems pour qui nous sommes devenus partenaires.

## 4. TRAVAUX APPARENTÉS

La plupart des modèles de composant font la distinction entre les propriétés fonctionnelles et les propriétés non-fonctionnelles. Pour notre modèle, nous avons également choisi de mettre en œuvre cette séparation des préoccupations avec le Composant Métier et le Processeur Élémentaire. Le concept de conteneur que nous utilisons pour le PE n'est pas présent dans tous les modèles. EJB [14] et CCM [9] proposent des conteneurs alors que Fractal [1] introduit la notion de contrôleur qui est une partie intégrante du composant chargée de gérer les propriétés non-fonctionnelles. Cependant les conteneurs de modèles comme EJB et CCM définissent un nombre figé de services qu'il n'est pas possible d'étendre. Dans nos travaux, nous avons besoin d'appliquer des politiques de communication telle la synchronisation entre flux, afin de rendre un service le plus proche possible des exigences de l'utilisateur.

Bon nombre de travaux portent sur les architectures logicielles de reconfiguration des réseaux de capteurs et sur la collaboration entre composants logiciels et matériels.

Les auteurs de [12] proposent une architecture pour configurer des réseaux de capteurs de façon rapide et dynamique. C'est une combinaison d'une architecture matérielle et d'une architecture logicielle qui permet l'échange d'information entre un réseau de capteur et d'autres applications. Un serveur gère les informations recueillies par les capteurs et les transmet vers les applications. Jusqu'à présent, lorsqu'une nouvelle fonctionnalité devait être chargée sur un capteur, il fallait mettre à jour le serveur pour qu'il charge le code nécessaire pour réaliser la translation des données brutes en données interprétables par les applications. [12] propose de détecter automatiquement la présence de nouveaux capteurs et d'en télécharger le composant logiciel permettant la translation des données. Ainsi il est possible de développer un serveur générique et réutilisable qui ne nécessite pas de mise à jour à chaque changement de fonctionnalité. Pour cela ils utilisent la technologie JavaBean pour décrire le réseau de capteur. Chaque capteur renferme deux JavaBeans : l'un contient les données spécifiques au capteur (Identifiant, localisation, type, etc.) et l'autre contient les méthodes pour accéder au capteur (translation des données, etc.). Le code occupe très peu de place dans la mémoire des capteurs, cependant la transmission d'un JavaBean peut demander une forte consommation d'énergie et réduire considérablement la durée de vie du capteur. Une alternative possible est de ne transmettre qu'une URL où le serveur peut ensuite télécharger le composant. Pour exporter les fonctions des capteurs et les rendre utilisables par les applications, un framework OSGi est installé sur le serveur. Ce framework offre une gestion aisée de la mobilité des capteurs par son mécanisme d'ajout et suppression de services.

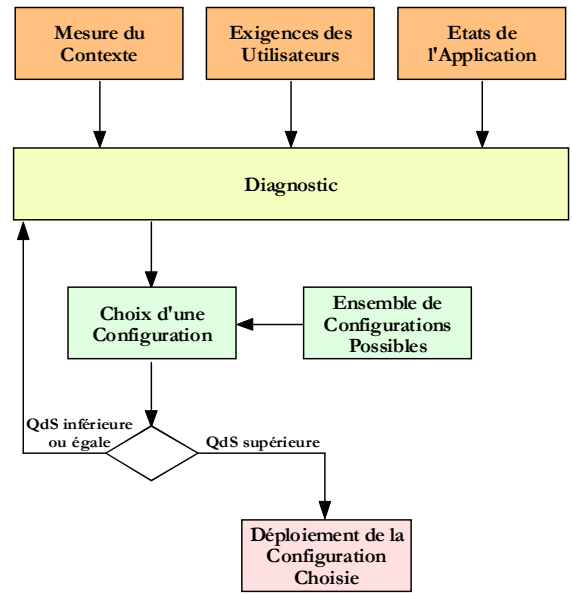

**Figure 5. Fonctionnement de la plateforme de reconfiguration**

L'architecture proposée dans [12] est intéressante par la flexibilité que procure le découplage entre les données spécifiques du capteur et son comportement. Dans nos recherches, nous avons également fait de choix de séparer deux aspects des composants : une partie fonctionnelle qui concerne le traitement des données et une partie non-fonctionnelle qui traite du transport de l'information et du contrôle de la partie fonctionnelle. Dans le même esprit de conservation de l'énergie, nous déportons des services de la plateforme et des parties des composants sur des hôtes moins contraints. Enfin, nous avons fait le choix d'une plateforme distribuée plutôt qu'un serveur centralisant les informations des capteurs et les transmettant aux applications. En effet l'évaluation de la qualité de service et le choix d'une nouvelle configuration devant être faits en temps réel, seule une méthode de calcul réparti peut y parvenir [7]. De plus, nous considérons les capteurs comme faisant partie intégrante des applications au même titre que les composants logiciels. Les capteurs et les composants logiciels collaborent directement pour réaliser le service rendu par l'application.

Les auteurs de [4] proposent un intergiciel évolutif selon les changements de l'environnement. Les capteurs doivent pouvoir adapter leur comportement en fonction des conditions environnementales. Ils doivent pouvoir également adapter leur mécanisme de communication ou encore leur protocole réseau. La reconfiguration des réseaux de capteurs est décrite selon deux dimensions : une dimension locale, au niveau du fonctionnement du capteur et une dimension distribuée, au niveau du fonctionnement du réseau de capteurs. Ils proposent deux architectures. La première est une architecture locale aux capteurs. Chaque capteur est un composant composite où les connexions entre les composants internes sont contrôlées par un ensemble de règles ordonnées par un configurateur. Lorsque survient un changement de contexte, le configurateur va engendrer une reconfiguration en ajoutant ou supprimant des composants internes et/ou des connexions. La deuxième architecture regroupe un ensemble d'architectures locales de même type, c'est-à-dire réalisant toutes la même fonctionnalité. Une reconfiguration au niveau réseau engendre une reconfiguration des architectures locales aux capteurs.

Tous les composants du réseau offrent la même fonctionnalité alors que nos applications sont le résultat de la collaboration de fonctionnalités différentes. L'architecture distribuée est similaire à notre plateforme distribuée sauf qu'elle ne se préoccupe que du fonctionnement du réseau et du calibrage des capteurs mais ne s'intéresse pas à l'ajout, la suppression et la modification des fonctionnalités des capteurs.

L'architecture proposée par [5][6] permet de modifier la fonctionnalité de chaque nœud du réseau de capteur en réponse à un changement de contexte environnemental ou applicatif. Chaque nœud renferme un composant *monitor*, un composant *configurator* et sa fonctionnalité. Les capteurs sont reliés à une base où sont hébergées les différentes configurations possibles, les exigences de QdS et le code à charger sur les nœuds correspondant à chaque configuration. Le composant m*onitor* mesure la QdS locale au capteur et la transmet à la base qui l'évalue. Il joue un rôle comparable à l'Unité de Contrôle des composants de notre architecture. Il faut toutefois remarquer que dans notre modèle, les PEs et les Conduits ayant une UC, notre évaluation de la QdS porte à la fois sur les traitements et sur les circulations d'informations. Le composant d'évaluation de la QdS de la base joue un rôle comparable au service *Supervision* que nous proposons. Le composant r*econfigurator* a la charge de reconfigurer la fonctionnalité locale au capteur sur ordre de la base. Il joue un rôle comparable au regroupement des services *Usine à Conteneur* et *Usine à Conduit* de notre architecture. Pour des raisons de conservation de l'énergie des capteurs, nous avons fait le choix de déporter les services *Usine à Conteneur* et *Usine à Conduit* ainsi que les UC des composants. De même, les reconfigurations dans [5][6] sont effectuées selon des configurations pré-établies comme il est fait dans [7]. Notre architecture ne se contente pas de reconfigurer les réseaux de capteurs, elle permet de reconfigurer des applications hétérogènes constituées de composants de nature différentes tels que les capteurs et les composants logiciels. De plus, elle permet de prendre en compte des contraintes liées au traitement de données multimédias comme la synchronisation.

## 5. CONCLUSION ET PERSPECTIVES

Jusqu'à présent, la plupart des applications utilisant des capteurs font une séparation entre le réseau de capteurs et les applications utilisant leurs informations. Nous proposons de dépasser cette distinction et de considérer les capteurs comme faisant partie intégrante des applications en proposant une architecture distribuée permettant de gérer et de reconfigurer à la fois les capteurs et les composants logiciels. De plus, les capteurs sont considérés comme de potentiels supports pour des composants logiciels. Ceci permet de proposer des organisations dans lesquelles les transferts d'informations qui sont de gros consommateurs d'énergie peuvent être minimisés en plaçant les composants de traitement sur ou au plus près de ceux qui produisent les informations à traiter. Ainsi, sur l'exemple de l'application de télésurveillance, le fait de pouvoir placer un composant de détection de mouvement sur un capteur vidéo utilisé en tant que détecteur de présence évite d'avoir à transmettre cette vidéo. De la même façon la possibilité d'effectuer un traitement sur un capteur servant de relais de transmission d'information peut permettre de ne transmettre que les informations strictement nécessaires.

L'architecture logicielle que nous proposons est constituée de quatre services : *Supervision*, *Usine à Conteneur*, *Usine à Conduit* et *Routage*. Le service *Supervision* surveille le niveau de QdS de l'application et réagit aux changements du contexte. En relation avec le service *Routage*, il contrôle les services *Usine à Conteneur* et *Usine à Conduit* pour ajouter et/ou supprimer des composants et des connexions adaptés aux besoins de l'application.

Peu de travaux sur les architectures pour les réseaux de capteurs relèvent le problème du déploiement et de l'hétérogénéité. Leurs applications sont constituées de dispositifs sensiblement identiques. Les applications visées par nos travaux sont constituées de composants aux contraintes différentes désignés comme hôtes légers et hôte fixes. Les composants et les connexions sont déployés différemment en fonction des contraintes de l'hôte. Nous répondons aux contraintes d'énergie et de mémoire des hôtes légers en déportant une partie des composants et des services de la plateforme sur un hôte fixe voisin. Le service *Routage* permet de faire face à la mobilité des capteurs et de déplacer les services déportés vers l'hôte fixe le plus proche.

Les travaux à court termes concernent le développement et le déploiement de l'architecture proposée sur un système composé de PCs et de capteurs Sun Spots. A moyens termes, ils concernent également la conception d'un système d'information unifié correspondant au fonctionnement de l'architecture c'est-à-dire à la description de tous les messages émis et reçus par la plateforme : création et destruction de conteneurs et de conduits, demande d'état de fonctionnement des composants et des conduits et demande d'état des routes.

## 6. REMERCIEMENTS

Ces travaux bénéficient du support l'ANR/CNRS dans le cadre du projet TCAP (Transport de flux vidéo sur réseaux de capteurs pour la surveillance à la demande) et de Sun Microsystems.